\documentclass[11pt]{article}
\pdfoutput=1
\usepackage{dcolumn}% Align table columns on decimal point
\usepackage{bm}% bold math

\usepackage[square,numbers,compress,sort]{natbib}
\usepackage{graphicx}% Include figure files
\usepackage{amssymb,amsmath}
\usepackage{multirow}
\usepackage{multicol}
\usepackage{nonfloat}
\usepackage{color,url}
\definecolor{vdrgreen}{rgb}{0.0, 0.7, 0.0}
\definecolor{vdrred}{rgb}{0.7, 0.0, 0.0}

\usepackage{tabu}

\usepackage{epsfig,psfrag,rotating,soul}
\usepackage{rotfloat}

\usepackage{slashed,xfrac}
\usepackage[makeroom]{cancel}
\usepackage{rotating}
\usepackage{booktabs}
\usepackage{color,url}
\usepackage[colorlinks=true
,urlcolor=blue
,anchorcolor=blue
,citecolor=blue
,filecolor=blue
,linkcolor=blue
,menucolor=blue
,linktocpage=true
,pdfproducer=medialab
,pdfa=true
]{hyperref}
\usepackage{mciteplus}
\usepackage{epsfig,psfrag,rotating,soul}
\usepackage{rotfloat}
\usepackage[normalem]{ulem}
\usepackage{framed,fancybox,xfrac,geometry}
\usepackage{hhline}
\usepackage{tabu}

\evensidemargin \oddsidemargin
\allowdisplaybreaks

\def\thefootnote{\fnsymbol{footnote}}

\renewcommand\eqref[1]{Eq.~(\ref{#1})}

\newcommand\figref[1]{Fig.~\ref{#1}}

\newcommand\tabref[1]{Table~\ref{#1}}

\newcommand{\mL}{{\mathcal L}}

\DeclareMathOperator{\Tr}{Tr}

\begin{document}
\thispagestyle{empty}

\begin{flushright}
IFT-UAM/CSIC-19-100\\
FTUAM-19-16\\
TUM-EFT-128-19\\
VBSCAN-PUB-07-19
\end{flushright}

\vspace{0.5cm}

\begin{center}

\begin{Large}
\textbf{\textsc{Dynamical vector resonances from the EChL in VBS at the LHC: the WW case}} 
\end{Large}

\vspace{1cm}

{\sc
R.L.~Delgado$^{1}$%
\footnote{\tt \href{mailto:rafael.delgado@tum.de}{rafael.delgado@tum.de}}%
, C.Garcia-Garcia$^{2}$%
\footnote{\tt \href{mailto:claudia.garcia@uam.es}{ claudia.garcia@uam.es}}%
, M.J.~Herrero$^{2}$%
\footnote{\tt \href{mailto:maria.herrero@uam.es}{maria.herrero@uam.es}}%
 }

\vspace*{.7cm}

{\sl 
$^1$ Physik-Department T30f, Technische Universit\"at M\"unchen (TUM),\\
Garching, Germany

\vspace*{0.1cm}

$^2$Departamento de F\'{\i}sica Te\'orica and Instituto de F\'{\i}sica Te\'orica, IFT-UAM/CSIC,\\
Universidad Aut\'onoma de Madrid, Cantoblanco, 28049 Madrid, Spain
 }

\end{center}

\vspace*{0.1cm}

\begin{abstract}
\noindent
In this work we study the phenomenology of the process $pp\to W^+W^-jj$ at the LHC, in the scenario of the resonant vector boson scattering subprocess  $W^+W^-\to W^+W^-$ which we describe within the effective field theory framework of the Electroweak Chiral Lagrangian. We assume a strongly interacting electroweak symmetry breaking sector in which dynamically generated resonances with masses in the TeV scale appear as poles in the Electroweak Chiral Lagrangian amplitudes unitarized with the Inverse Amplitude Method. The relevant resonance here, $V_0$,  is the neutral component of the triplet of vector resonances which are known to emerge dynamically at the TeV scale for specific values of the Electroweak Chiral Lagrangian parameters. With the aim of studying the production and possible observation of $V^0$ at the LHC, via the resonant $W^+W^-\to W^+W^-$ scattering, a MadGraph~5 UFO model has been developed employing a phenomenological Proca Lagrangian as a practical tool to mimic the correct $V^0$ properties that are predicted with the Inverse Amplitude Method. We choose to study the fully hadronic decay channel of the final gauge bosons $W W\to J(jj)J(jj)$ since it leads to larger event rates and because in the alternative leptonic decay channels the presence of neutrinos complicates the reconstruction of the resonance properties. In this context, the 2 boosted jets from the $W$ hadronic decays, $jj$, are detected as a single fat jet, $J$, due to their extreme collinearity. We perform a dedicated analysis of the sensitivity to these vector resonances $V^0$ with masses between 1.5 and 2.5 TeV at the LHC with $\sqrt{s}=13$ TeV and the planned high luminosity of 3000 ${\rm fb}^{-1}$, paying special attention to the study of efficient cuts to extract the resonant vector boson fusion signal from the QCD background, which clearly represents the main challenge of this search.
\end{abstract}

\def\thefootnote{\arabic{footnote}}
\setcounter{page}{0}
\setcounter{footnote}{0}

\newpage

\section{Introduction}
\label{intro}
%%%%%%%%%%%%%%%%%%%%%%%%%%%%

The emergence of heavy resonances in Vector Boson Scattering (VBS) with external electroweak (EW) gauge bosons, $W^+$, $W^-$ and $Z$ would be undoubtedly a remarkable signal of new physics beyond the Standard Model (SM) of fundamental interactions. If these new resonances have masses in the few TeV energy domain, the LHC is then the  proper collider to look for them. In the case in which these resonances couple dominantly to EW gauge bosons, and not to fermions, it is clear that the VBS kind of subprocesses plays the most relevant role in the search for these emergent resonances. In particular, these resonant states could show up via the study of events at the LHC with two electroweak bosons and two tagged jets with VBS configuration, i.e., with large invariant mass $M_{jj}$ and large rapidity separation $\Delta \eta_{jj}$. This type of events will provide the cleanest and more efficient window to look for them and to study their properties, given the available energies at this proton-proton collider and also given the forthcoming LHC period with high planned luminosities from $300\,{\rm fb}^{-1}$ up to $3000\,{\rm fb}^{-1}$.

Clearly, the natural framework for these resonances emerging at VBS is provided by the class of theories where the self interactions among the longitudinal gauge bosons $W_L$ and $Z_L$ are assumed to become strong at the TeV scale, a hypothesis which is interestingly suggested by the peculiar  growing behaviour with energy of the VBS cross section for these polarization modes in presence of anomalous (non-standard) gauge boson self-couplings. As for the particular theory describing this new strongly interacting dynamics in the electroweak symmetry breaking (EWSB) sector and the associated emergent resonances, we will not assume any specific underlying fundamental model, but instead we will work within the effective field theory framework provided by the Electroweak Chiral Lagrangian (EChL), which is the proper tool for a generic model independent prediction. The EChL is simply based on the electroweak chiral symmetry breaking pattern $SU(2)_L \times SU(2)_R \to SU(2)_{L+R}$, within a non-linear realization,  and involves the minimal set of three Goldstone modes providing the longitudinal components of the EW gauge bosons and, therefore, their masses. The earliest version of the EChL was built through the eighties and nineties~\cite{Appelquist:1980vg,Longhitano:1980iz,Longhitano:1980tm,Chanowitz:1985hj,Cheyette:1987jf,Dobado:1989ax,Dobado:1989ue,Dobado:1990jy,Dobado:1990am,Dobado:1990zh,Espriu:1991vm,Feruglio:1992wf,Dobado:1995qy,Dobado:1999xb}, and it was renewed and completed years later to include the Higgs particle after its discovery~\cite{Alonso:2012px,Buchalla:2013rka,Espriu:2012ih,Delgado:2013loa,Delgado:2013hxa,Brivio:2013pma,Espriu:2013fia,Espriu:2014jya,Delgado:2014jda,Buchalla:2015qju,Arnan:2015csa,Buchalla:2017jlu}. We will focus here on the bosonic part of the EChL and ignore the fermionic sector, since we are mostly interested in the phenomenology associated to the case where the self-interactions among the longitudinal EW gauge bosons are the key to the new physics.

Within the EChL framework the natural scale for the expected resonances to appear is related to $4\pi v \sim$ 3 TeV, with $v=246\,{\rm GeV}$, which is the typical parameter with dimension of energy controlling the perturbative expansion in derivatives, or powers of energy, of this chiral effective field theory. Thus, the few TeV energy range is a well motivated mass assumption for these resonances, and the VBS processes the best place to look for them at the LHC. Within this context, the resonances emerge as poles in the total resumed EChL amplitude, taking into account the subsequent re-scattering of the EW gauge bosons via VBS type of diagrams, and adding their contributions to the total cross section, which is important in the case of strong interactions. This motivates the name ``dynamical resonances''. This qualitative description of the dynamical resonances emerging in VBS is inspired in the well known case of low energy QCD where, for instance, the $\rho$ particle is an emergent resonance in pion-pion scattering and its properties  are efficiently studied by means of those scattering processes within Chiral Perturbation Theory.

Here we use the Inverse Amplitude Method (IAM)\footnote{For a review, see, for instance,~\cite{Oller:1997ng,GomezNicola:2001as}.} to deal with this resummation process in VBS and to get unitary predictions at the same time.  The IAM unitarized partial wave amplitudes that are predicted from the EChL for a given set of the EChL parameters (also called coefficients in the effective operators of the Chiral expansion) have poles for particular values of these coefficients. These fix the physical properties of the resonances, like the mass, the width and the couplings to the EW gauge bosons. We follow here very closely the IAM approach of~\cite{Delgado:2017cls}, where this method was applied to the study of $WZjj$ events at the LHC via the $WZ \to WZ$ channel. In particular we take from this reference the description of the isotriplet vector resonances, $V^0$, $V^+$ and $V^-$, that appear as poles in the $IJ=11$ case, with $I$ and $J$ being the weak isospin and total angular momentum, respectively. In contrast to this reference, where the $W^\pm Z$ channels give access to the charged resonances $V^{\pm}$, here we focus in a different VBS subprocess, concretely, $W^+W^- \to W^+W^-$ which gives access to the neutral vector resonance $V^0$ via the  s-channel. We center our analysis on the hadronic decays of the final $W$s, more concretely in the kinematical regime in which the hadronic decay products of the $W$s are identified as a single, large radius jet, also in contrast to \cite{Delgado:2017cls}, where the focus was set on the leptonic decays. This allows for larger signal rates and for a better reconstruction of the resonance properties, since, in the alternative leptonic decays of the final $W$s, the large amount of missing energy in the final state complicates this task. However, this channel suffers from quite sizable backgrounds, especially regarding the one coming from QCD events. The biggest effort in this work is devoted to the optimization of the analysis of the $W$ tagging techniques with fat jets for the observation of this emergent $V^0$ resonance for the case of $W^+W^-jj$ events in the difficult hadronic channel. All in all, our analysis aims to explore the sensitivity to the neutral vector resonances $V^0$ with masses between 1.5 and 2.5 TeV at the LHC with $\sqrt{s}=13$~TeV and the planned high luminosity of $3000\,{\rm fb}^{-1}$, paying special attention to the study of efficient cuts to extract the resonant  signal from the QCD background in $pp \to JJjj$ events with VBS configuration, which clearly represents the main challenge of this search.

The paper is organized as follows: in Section \ref{model}, we summarize the theoretical framework we rely upon, including the relevant terms of the Electroweak Chiral Lagrangian and the effective Proca Lagrangian, as well as a revision of the main properties of the unitarized partial waves by means of the Inverse Amplitude Method. The analytical prediction of the scattering amplitude for $W^+W^- \to W^+W^-$ in that theoretical context is also presented in this section in a convenient form for a Monte Carlo implementation based on interaction vertices like MadGraph. In Section \ref{results} we present the numerical results of our study of the resonant $pp \to W(J) W(J) jj$ events at the LHC. Section \ref{conclusions} summarizes our main conclusions. 
      
%%%%%%%%%%%%%%%%%%%%%%%%%%%%%%%%%%

\section{Theoretical framework}
\label{model}
Here we work within the Electroweak Chiral Lagrangian framework,  whose main properties relevant for the computation of VBS observables are encoded in the following terms: 
%%%%%%%%%%%%%%%%%%%%%%% EChL %%%%%%%%%%
\begin{align}
    &\mL_{\rm EChL} = \mL_2 +\mL_4+\dots\\
   & \mL_2 = \frac{v^2}{4}\mathcal{F}(H)\Tr(D_\mu U^\dagger D^\mu U) + \frac{1}{2}\partial_\mu  H\partial^\mu H +\dots  \\
    &\mL_4 = a_4[\Tr({\cal V}_\mu  {\cal V}_\nu)][\Tr({\cal V}^\mu 
    {\cal V}^\nu)] \nonumber +a_5[\Tr({\cal V}_\mu {\cal V}^\mu)][\Tr({\cal V}_\nu {\cal V}^\nu)]+\dots 
\end{align}
where
\begin{align} 
&U =\exp\left(i\omega^a\tau^a/v\right), \,\, {\cal V}_\mu = (D_\mu U)U^\dagger, \,\,\mathcal{F}(H)=1+2a\frac{H}{v}+b\left(\frac{H}{v}\right)^2+\dots\nonumber\\
&D_\mu U = \partial_\mu U + i g (\vec{W}_\mu \vec{\tau}/2) U -i g'U (B_\mu \tau^3/2)\,.
 \end{align}
%%%%%%%%%%%%%%%%%%%Unitarity and the IAM %%%%%%%%%%
An extended version of this Lagrangian as well as the counting rules followed here can be found in~\cite{Delgado:2017cls}.

The amplitudes for VBS that are predicted from the previous $\mL_{\rm EChL}$ are well known, as well as the fact that, due to the typical growing with energy of these amplitudes, they may cross the unitarity limit at energies available at the LHC. The need of an unitarization method in order to provide realistic predictions at the LHC leads to results for the VBS mediated processes which are, in general, unitarization model dependent (for a recent review, see for instance \cite{Garcia-Garcia:2019oig}). 
We choose here one of the most commonly used unitarization methods for the partial waves, the IAM, which has the advantage over other methods of being able to generate dynamically the vector resonances that we are interested in. These are named dynamical resonances because they emerge in VBS as a consequence of the assumed strong dynamics involved in the iterative EW gauge bosons re-scattering, which  is dominated mainly by the longitudinal polarization degrees.  

In terms of fixed isospin $I$ and angular momentum $J$, the IAM partial waves are given by: 
\begin{align}
   a^{\rm IAM}_{IJ} = \frac{\big(a_{IJ}^{(0)}\big)^2}
 {a_{IJ}^{(0)}-a_{IJ}^{(1)}} \, ,
\label{IAMdef}
\end{align}
where $a_{IJ}^{(0)}$ is the prediction from $\mL_{\rm EChL}$ at Leading Order (LO), i.e., from $\mL_2$ at tree level, and $a_{IJ}^{(1)}$ is the Next to Leading Order (NLO) prediction, i.e., the one coming from $\mL_4$ at tree level and from $\mL_2$ at one-loop. The resonant behavior of the previous IAM amplitudes occurs for a fixed set of values of the EChL parameters at the energies, $\sqrt{s}$, where \eqref{IAMdef} has a pole. These particular EChL parameters then provide the values of the physical properties characterizing the emergent dynamical resonances. In particular, for the $IJ=11$ channel the pole at $s_{\rm pole}=\left(M_{V}-\frac{i}{2}\Gamma_{V}\right)^2$ gives the values of, $M_V$ and $\Gamma_{V}$ for the iso-vector resonances in terms of the considered EChL parameters, namely, $a$, $b$, $a_4$ and $a_5$ in our simplified version here. In fact, for the numerical estimates in the present work, we will further simplify these to just three parameters, $a_4$, $a_5$ and $a$, and set $b=a^2$, being the latter condition well motivated in a wide variety of models. 
For example, the particular relation, $b=a^2$, between the two EChL coefficients $a$ and $b$ is present, first of all, in the SM itself, where $b=a^2=1$. The same condition $b=a^2$ also appears in the so called dilaton models (see, for instance refs.~\cite{Goldberger:2008zz,Fan:2008jk,Vecchi:2010gj}), where a light dilaton, identified with the Higgs particle,  arises from the spontaneous breaking of the scale invariance of the strong sector of the model. In that case,  conformal invariance requires $b=a^2$. 
We have taken this relation, however,  just as a simplification hypothesis in our work, in order to reduce the number of input parameters in our forthcoming Monte Carlo analysis of signal versus background. We have explicitly checked  that the resonance properties (the only ones affected by the $b$ parameter since it does not enter at the tree level in the studied scattering) do not change significantly if one assumes $b\neq a^2$, as it is mentioned in ref.~\cite{Delgado:2017cls}.

In order to study the production of these dynamical vector resonances at the LHC by means of a Monte Carlo like MadGraph~\cite{Alwall:2014hca,Frederix:2018nkq}, where the needed input files are not the scattering amplitudes but the interaction vertices themselves, or equivalently the interaction Lagrangian, we use the Proca Lagrangian as a phenomenological approach to mimic the IAM vector resonances. In this approach we follow closely ~\cite{Delgado:2017cls}. This Proca Lagrangian is the simplest Lagrangian description for vector resonances that shares the chiral and gauge symmetries of the EChL, and is given, in its simplest form by:
\begin{align} 
\mL_V =& -\frac{1}{4}{\rm Tr}( {\hat V}_{\mu\nu} {\hat V}^{\mu\nu}) +
 \frac{1}{2} M_V^2 {\rm Tr}( {\hat V}_\mu {\hat V}^\mu )+ \frac{i g_V}{2\sqrt{2} } {\rm Tr}( {\hat V}_{\mu\nu} \, [u^\mu,u^\nu ] )\,.
\label{Proca}
\end{align}
The isotriplet vector resonances, $V^{\pm}$ and $V^0$, are implemented via the ${\hat V}_\mu$ fields: 
\begin{align}
{\hat V}_\mu &=
\left(\begin{array}{cc}
\frac{V^0_\mu }{\sqrt{2}} & V^+_\mu \\ V^-_\mu  &-\frac{V_\mu^0}{\sqrt{2}}
\end{array}\right)\, ,
\end{align}
where the three states have mass $M_V$, and couplings to the EW gauge bosons set by the parameter $g_V$. The definitions of the remaining fields and tensors involved in \eqref{Proca} are as follows:
\begin{align}
&{\hat V}_{\mu\nu} = \nabla_\mu {\hat V}_\nu -\nabla_\nu {\hat V}_\mu\, , 
\nonumber \\
&u_\mu  =    \,
i\, u\, \Big(D_\mu U\Big)^\dagger u \,\,, {\rm with}\,\, u^2=U \, , \nonumber \\
&\nabla_\mu X \, =\, \partial_\mu X \, +\, [\Gamma_\mu , X ] \,\,, {\rm with}\,\, \Gamma_\mu =
\frac{1}{2} \Big(\Gamma_\mu^{L} +\Gamma_\mu^{R}\Big)\, , 
 \nonumber \\
&\Gamma_\mu^{L} = u^\dagger \left(\partial_\mu + i\,\frac{g}{2} \vec{\tau}\vec{W}_\mu
\right) u^{\phantom{\dagger}} 
\, , \quad\;
\Gamma_\mu^{R} = u^{\phantom{\dagger}}  \left(\partial_\mu + i\, \frac{g'}{2} \tau^3 B_\mu\right)u^\dagger 
\, .
\end{align}
As it has been proven in ~\cite{Delgado:2017cls}, $\mL_V$  mimics quite well the properties of the vector resonances that are dynamically generated with the IAM from the EChL, once the model parameters, $M_V$ and $g_V$ are set properly in terms of the EChL parameters, concretely $a_4$, $a_5$ and $a$ (recall that $b$ is simplified to $b=a^2$).   
%%%%%%%%%%%%%%

\begin{figure}[t!]
\begin{center}
\includegraphics[width=0.9\textwidth]{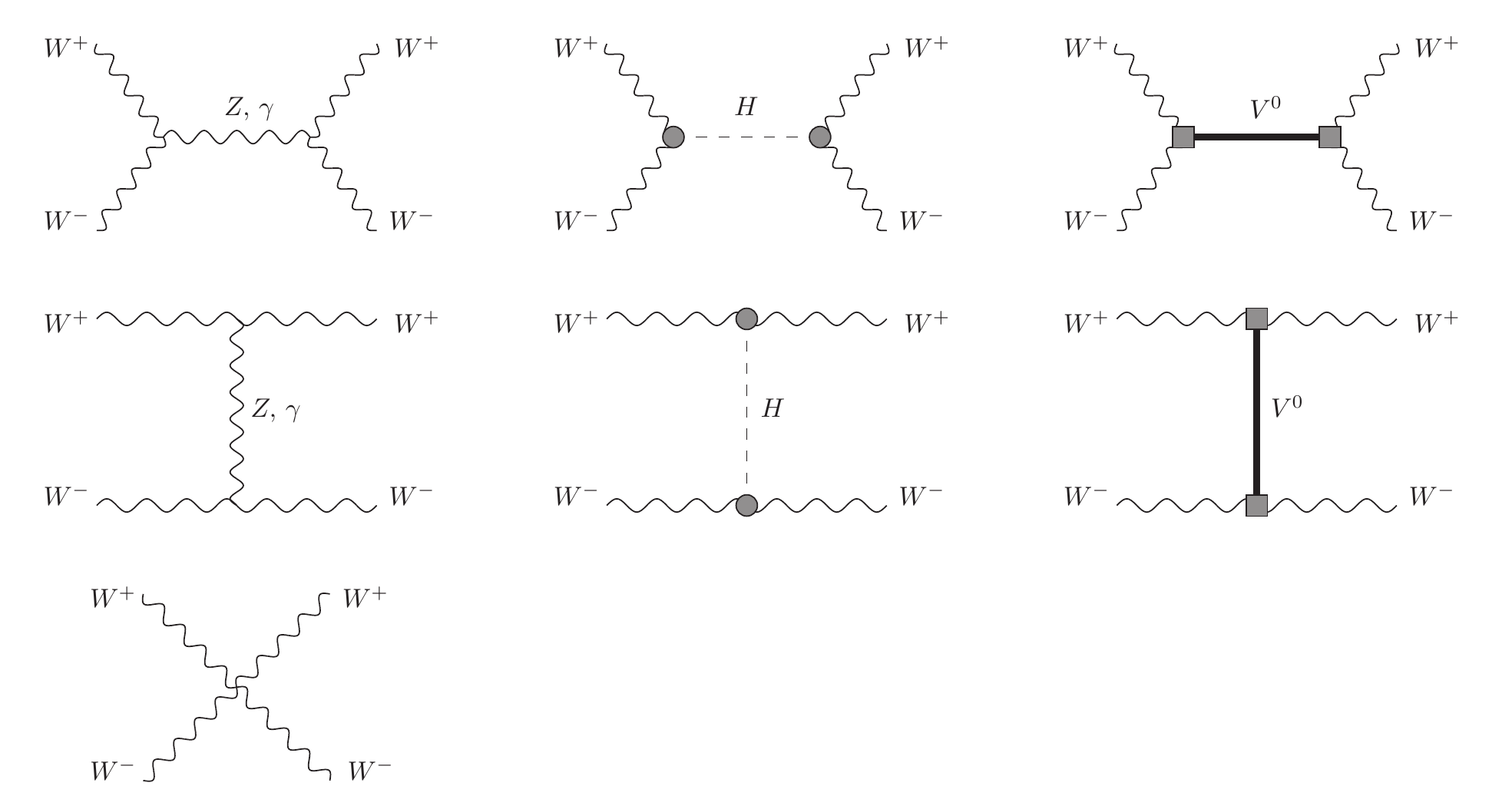}
\caption{Feynman diagrams contributing to the process $W^+W^-\to W^+W^-$ in the Unitary gauge  within our effective  Lagrangian formalism. Gray circles (middle diagrams) represent the $V_{W^+W^-H}$ vertex given in \eqref{vertexWWH} in terms of the EChL parameter $a$ and gray squares (right diagrams) represent the $V_{W^+W^-V^0}$ vertex in \eqref{vertexWWV} in terms of the $g_V$ Proca coupling.}
\label{FD}
\end{center}
\end{figure}

From the above Proca Lagrangian, $\mL_V$, and the previous ${\cal L}_2$ in the EChL one can derive easily all the relevant Feynman rules for the scattering of interest here $W^+W^- \to W^+W^-$. The relevant interaction vertices which are either new or modified with respect to the SM ones are: 
\begin{itemize}
\item[1)] the interaction vertex between one vector resonance and two EW gauge bosons, given by:
\begin{align}
V_{W^+_\mu W^-_\nu V^0_\rho} &=i g^2 g_V(g_{\mu \rho} k_\nu - g_{\nu \rho} k_\mu)/2,
\label{vertexWWV}
\end{align}
where $k$ is the momentum of the vector resonance taken as incoming to the interaction vertex, and 
\item[2)] the interaction vertex between one Higgs boson and two EW gauge bosons, given by:
\begin{align}
V_{W^+_\mu W^-_\nu H}&=iagM_W g_{\mu\nu},
\label{vertexWWH}
\end{align}
where $a$ is the EChL parameter introduced above.
\end{itemize}
The total set of Feynman diagrams contributing to the $W^+W^- \to W^+W^-$ scattering is collected in Figure \ref{FD}, where the above commented interaction vertices $V_{W^+W^-V^0}$ and $V_{W^+W^-H}$ are represented by a shadowed square and a shadowed circle respectively. For the practical computation of the full process $pp \to W^+W^-jj$ with MadGraph we have implemented an UFO model with the complete Lagrangian, involving the two relevant pieces $\mL_2$ and $\mL_V$. It is worth mentioning that the full set of diagrams involved in $pp \to W^+W^-jj$, which are generated in MadGraph, include many other diagrams in addition to the subset of diagrams with VBS configuration, so all of them have been taken into account in our numerical computation of the LHC events presented in the next section.
 
%%%%%%%%%%%%%%%%%%%%%5

\begin{table}[t!]
\centering
\vspace{.2cm}
%\begin{tabular}{|c p{\mylength}|c p{\mylength}|c p{\mylength}|c p{\mylength}|c p{\mylength}|c p{\mylength}|c p{\mylength}|}
\begin{tabu} to 0.8\textwidth {X[c]X[c]X[c]X[c]X[c]X[c]X[c]}
%\begin{tabular}{ |c|c|c|c|c|c|c| }
%\rowcolor{gray! 50}
\toprule
\toprule
{\footnotesize {\bf BP}} & {\footnotesize {\bf $M_V $}}  & {\footnotesize  {\bf $\Gamma_V $}}  & {\footnotesize {\bf $g_V(M_V^2)$}}  & {\footnotesize {\bf $a$}}  & {\footnotesize {\bf $a_4 \cdot 10^{4}$}}  & {\footnotesize {\bf $a_5\cdot 10^{4}$}}
\\
\midrule
BP1  & $\quad 1476 \quad $ & $\quad 14 \quad $ & $ \quad 0.033  \quad $ & $ \quad 1 \quad $ & $ \quad 3.5 \quad $ & $ \quad -3 \quad $
\\[3pt] 
BP2  & $\quad 2039 \quad $ & $\quad 21 \quad $ & $ \quad 0.018  \quad $ & $ \quad 1 \quad $ & $ \quad 1 \quad $ & $ \quad -1 \quad $
\\[3pt] 
BP3  & $\quad 2472 \quad $ & $\quad 27 \quad $ & $ \quad 0.013  \quad $ & $ \quad 1 \quad $ & $ \quad 0.5 \quad $ & $ \quad -0.5 \quad $
\\[3pt]
BP1' & $\quad 1479 \quad $ & $\quad 42 \quad $ & $ \quad 0.058  \quad $ & $ \quad 0.9 \quad $ & $ \quad 9.5 \quad $ & $ \quad -6.5 \quad $
\\[3pt]
BP2'  & $\quad 1980 \quad $ & $\quad 97 \quad $ & $ \quad 0.042  \quad $ & $ \quad 0.9 \quad $ & $ \quad 5.5 \quad $ & $ \quad -2.5\quad $
\\[3pt]
BP3'  & $\quad 2480 \quad $ & $\quad 183 \quad $ & $ \quad 0.033  \quad $ & $ \quad 0.9 \quad $ & $ \quad 4\quad $ & $ \quad -1 \quad $
\\
\bottomrule\bottomrule
\end{tabu}
\vspace{0.4cm}
\caption{\small Selected benchmark points (BPs) of dynamically generated vector resonances taken from 
\cite{Delgado:2017cls}. The mass, $M_V$, in GeV; width, $\Gamma_V$, in GeV; and coupling to gauge bosons, $g_V(M_V^ 2)$, as well as the relevant chiral parameters, $a$, $a_4$ and $a_5$ are given for each of them. In all these points we have set $b=a^ 2$. 
}
\label{tablaBMP}
\end{table}
%%%%%%%%%%%%%%%%%%%%%%%
 
The analytical prediction for $W^+W^- \to W^+W^-$ can be written in a very simple form by means of the following 4-point effective function. Using a short-hand notation in the following subscripts of  these functions given by $4W= W^+_\mu (k_1) W^-_\nu (k_2) W^+_\sigma (k_3) W^-_\lambda (k_4)$, we get:
\begin{align}
-i\,\Gamma_{4W}^{\rm eff} & =
-i\,\Gamma_{4W}^{\mL_2} 
-i\,\Gamma_{4W}^{\mL_V} =-i\,\Gamma_{4W}^{\rm SM} 
-i\,\Gamma_{4W}^{ (a-1)} 
-i\,\Gamma_{4W}^{\mL_V}\,.
\label{fpfUFO}
\end{align}
Here $\Gamma^{\rm SM}$ is the contribution from the SM, $\Gamma^{(a-1)}$ denotes the new effects introduced by  $\mL_2$ with $a\neq1$ with respect to the SM, and $\Gamma^{\mL_V}$ accounts for the new contributions from the dynamically generated resonance. The decomposition defined in \eqref{fpfUFO} turns out to be very convenient to introduce our model in MadGraph, as one can use the SM default model as the basic tool to build the UFO. In this way, we just add up to the SM model files the $\Gamma^{(a-1)}$ and $\Gamma^{\mL_V}$ as four point functions given by:
\begin{align}
-i\Gamma_{4W}^{(a-1)} =& 
	-g^2\frac{m_W^2}{t-m_H^2}(a^2-1) g_{\mu\sigma} g_{\nu\lambda} -g^2\frac{m_W^2}{s-m_H^2}(a^2-1) g_{\mu\nu} g_{\sigma\lambda}  \nonumber\\
-i\Gamma_{4W}^{\mL_V} =&  
	\frac{g^4}{4}\bigg[\frac{g_V^2(s)}{s-M_V^2+iM_V\Gamma_V} (h_\nu h_\lambda g_{\mu\sigma}-h_\nu h_\sigma g_{\mu\lambda}-h_\mu h_\lambda g_{\nu\sigma}+h_\mu h_\sigma g_{\nu\lambda}) \nonumber\\
	&+\frac{g_V^2(t)}{t-M_V^2}(l_\nu l_\lambda g_{\mu\sigma}-l_\lambda h_\sigma g_{\mu\nu} -l_\mu l_\nu g_{\lambda\sigma} + l_\mu l_\sigma g_{\nu\lambda})\bigg]\,,
\end{align}
where $h=k_1+k_2$ and $l=k_1-k_3$. 

In the above expressions an energy dependent coupling has been introduced which, following~\cite{Delgado:2017cls}, is defined in terms of the Proca coupling $g_V$ as:
\begin{align}
g_V^2(z)&=g_V^2(M_V^2) \frac{M_V^2}{z} \,\,\, {\rm for} \,\, s< M_V^2 \nonumber, \\
g_V^2(z)&=g_V^2(M_V^2) \frac{M_V^4}{z^2} \,\,\, {\rm for} \,\, s> M_V^2\,,
\label{gvenergytu}
\end{align}

with $z=s,t$ corresponding to the $s,t$ channels, respectively, in which the resonance $V^0$ is propagating in the present case of  $W^+W^-$ scattering. These non-local interactions are needed in order to ensure unitary predictions, since the Proca Lagrangian itself, i.e., with a constant $g_V$, leads to a violation of unitarity above the resonance mass. For more details on the model see \cite{Delgado:2017cls}.

In \tabref{tablaBMP} we collect a number of selected benchmark points (BP) taken from \cite{Delgado:2017cls}; namely, some specific sets of values for the relevant parameters $a$, $a_4$ and $a_5$ that yield to dynamically generated vector resonances emerging in the $IJ=11$ channel with masses around $1.5$, $2$ and $2.5\,{\rm TeV}$. These particular mass values for the vector resonances, belonging to the interval $1-3\,{\rm TeV}$, have been chosen on purpose as illustrative examples of the a priori expected reachable masses at the LHC. In the following section we will use these benchmark points to predict the visibility of vector resonances that may exist in the $IJ=11$ channel, and therefore resonate in the process $WW \to WW$ at the LHC.

%%%%%%%%%%%%%%%%%%%%%%%%%%%5

%
\section{Numerical Results}
\label{results}
In this section we present the numerical results of the $pp \to W^+W^-jj$ events at the LHC computed within our model for the isotriplet vector resonances  described in the previous section, selecting exclusively the hadronic decays of the final $W$ gauge bosons. In the present case, as we have said, the relevant resonance is the neutral vector resonance $V^0$.  This is in contrast to the previous study in~\cite{Delgado:2017cls}, where the same theoretical framework was considered to describe the isotriplet vector resonances, but the focus was set mainly on the EW gauge boson leptonic channels of the $pp \to WZjj$ events at the LHC, where the relevant vector resonances are the charged ones $V^{\pm}$. Thus, the present analysis is somehow complementary to the previous work in~\cite{Delgado:2017cls}.  In all this section we set the LHC energy to $\sqrt{s}=13$ TeV, and make predictions for all our signal benchmark scenarios defined by the six selected points BP1, BP2, BP3, BP1', BP2', BP3', collected in \tabref{tablaBMP}.  

In order to get a rough estimate of the signal rates at the LHC, and to learn about the main features of our signal events for the BPs selected points, we first analyze them at the naive parton level. We compute the rates for $pp \to W^+W^-jj$ events, with $jj$ denoting quarks, before considering any showering or jet reconstruction algorithm. Then, we apply the suppression factors from the two EW gauge boson hadronic decays given by $({\rm BR}(W \to {\rm hadrons}))^2\sim0.45$. In this simple computation of the cross sections for the $pp \to W^+W^-jj$ process we also wish to compare the signal rates with the main background rates which at this parton level are: 1) SM EW background, with amplitude of ${\cal O}(\alpha^2)$, 2) SM mixed QCDEW background, with amplitude of ${\cal O}(\alpha \alpha_S)$ and 3) top-antitop production from QCD followed by the top (antitop) decay into $b W^+$ $\left({\bar b} W^-\right)$, in which the final bottom and antibottom jets are misidentified as light quark jets $j$. In this latter case we assume a suppression factor due to this misidentification of $bb$ as $jj$ within the range from $(0.2)^2$ to $(0.3)^2$ corresponding to the often used b-jet tagging efficiency of 80\% to 70\%. Since we are interested in events with VBS scattering configuration and also within the large invariant mass region of the gauge boson pair, $M_{WW}$, for this parton level computation we have applied in addition to the basic cuts that ensure particle detection, $p_T^j>20$~GeV, $\Delta R_{jj}>0.4$, $\lvert\eta_W\rvert<2$,  $p_T^W>20$~GeV, also the usual VBS cuts given by:
%%%%%%%%%%%%%%%%%%%%%%%%%%%%%%%%%%
\begin{figure}[t!]
\begin{center}
\includegraphics[width=0.49\textwidth]{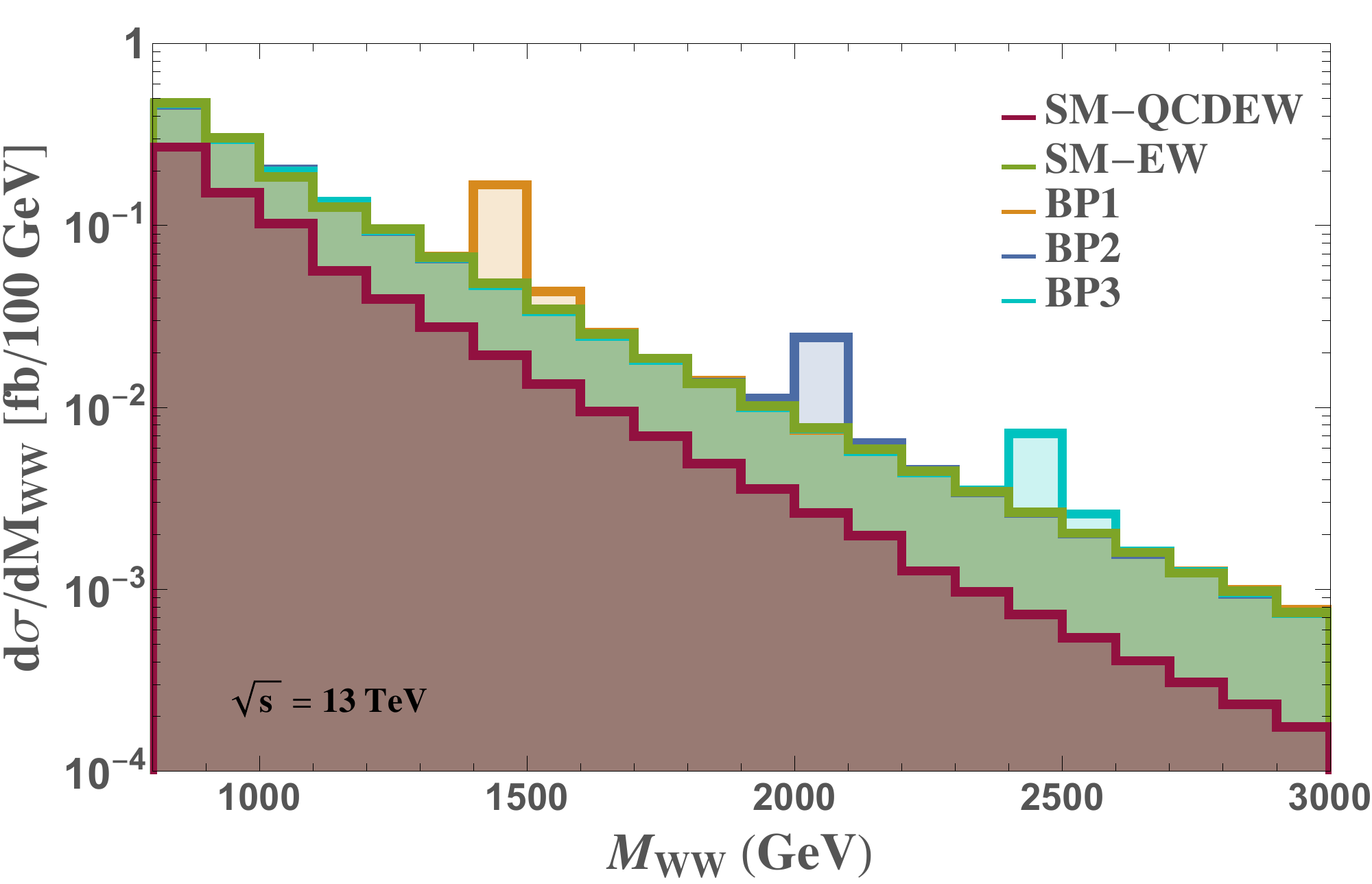}
\includegraphics[width=0.49\textwidth]{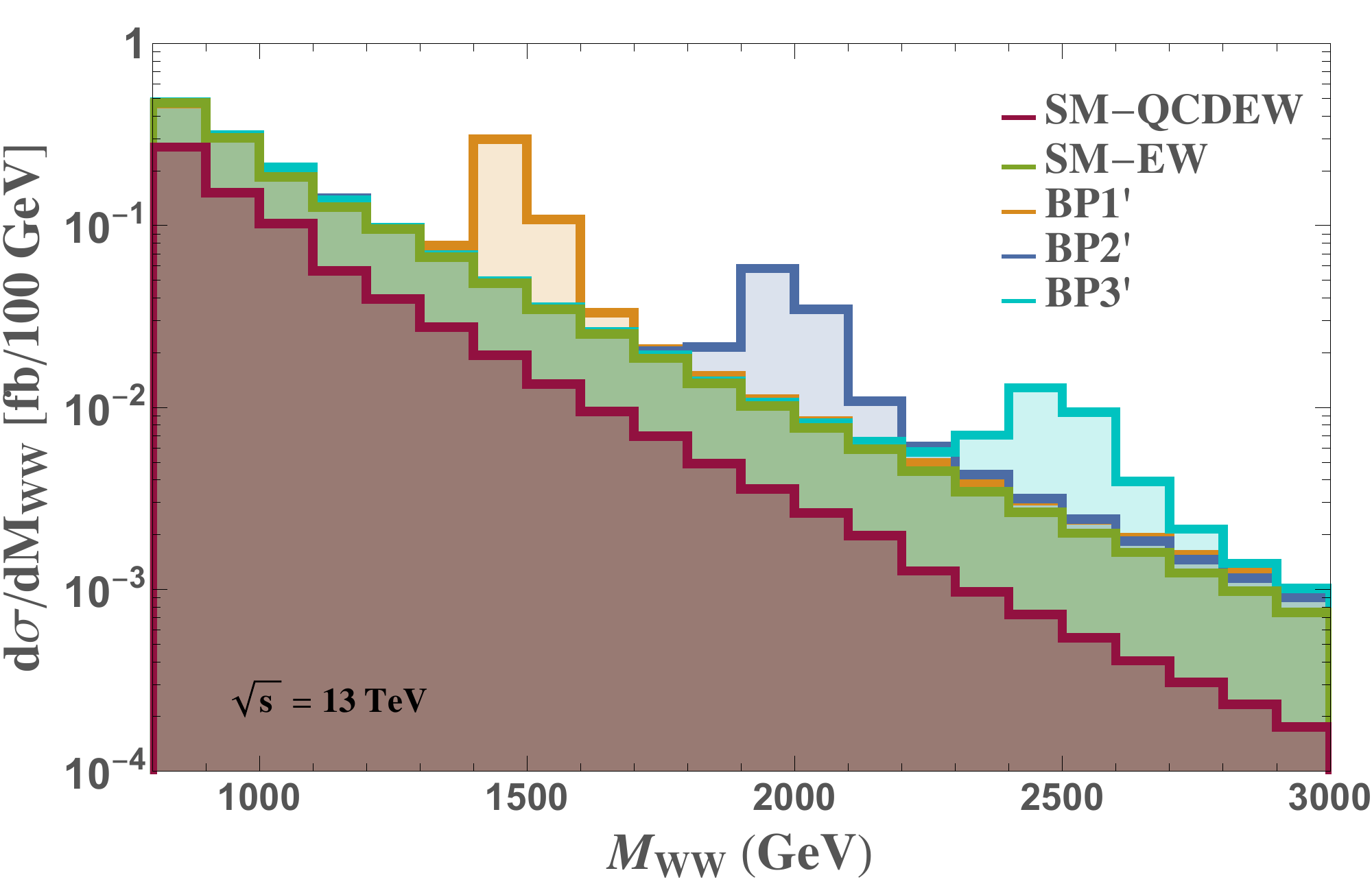}
\caption{Predictions of the cross sections times the branching ratios of the $W$ bosons to hadrons, $\sigma(pp \to W^+W^-jj)\times({\rm BR}(W\to {\rm hadrons}))^2$, distributions with the invariant mass of the EW gauge boson pair, $M_{WW}$, at the parton level, using MadGraph. The rates for the EW background (SM-EW), the mixed QCD-QED background (SM-QCDEW) and the selected signal scenarios for the vector resonances given by the BP's defined in \tabref{tablaBMP} are included. The cuts in \eqref{VBScuts} have been applied.}
\label{fig:partonlevel}
\end{center}
\end{figure}
%%%%%%%%%%%%%%%%%%%%%%%%%%%%%%%%%%%%%%%%%%
\begin{gather}
 2 < \lvert\eta_{j_1,j_2}\rvert < 5, \nonumber \\
 \eta_{j_1}\cdot\eta_{j_2} < 0,\nonumber \\\
  M_{j_1j_2}>500~{\rm GeV},
\label{VBScuts}
\end{gather}
where $j_1$ and $j_2$ here refer to the two final quarks produced together with the two $W$s. With these specific cuts, the kinematical configurations of the final quark-jets are typically within two opposite-sign pseudorapidity cones  with large differences in pseudorapidity and with large invariant mass of the jet pair, whereas the final EW gauge bosons are produced mainly in the central region. All these parton level predictions are obtained with MadGraph, and the results are collected in \figref{fig:partonlevel} and in \tabref{parton}. \figref{fig:partonlevel} shows the distributions in the invariant mass of the $WW$ pair and \tabref{parton} summarizes the total cross sections in the invariant mass region of the $WW$ pair of our interest, i.e., summing events over the interval $800\,{\rm GeV}<M_{WW}<3000\,{\rm GeV}$. As we can see in \figref{fig:partonlevel} the six studied resonances emerge clearly above the prediction from the SM continuum, which is in turn clearly dominated by the EW background. The other studied backgrounds, the mixed QCD-EW and the top-antitop are clearly subdominant for these specific configurations. In fact, we have checked explicitly that the main responsible for this strong suppression of the  mixed QCD-EW and the top-antitop backgrounds are the VBS cuts of \eqref{VBScuts}. In particular, the top-antitop background is reduced by a factor of about $10^{-3}$ when applying these VBS cuts in the selected region of large $M_{WW}$.   

Although these results at the parton level are encouraging, the real challenge is to deal with the difficult task of reconstructing the $W$s from their hadronic decay products. This is the issue that we confront next. 
Concretely, we are going to explore the prospectives for the fully hadronic decay channel,
\begin{equation}\label{mc_process}
 pp\to W^+W^-jj,~ W^\pm\to J(jj),
\end{equation}
where the 2 jets ($jj$) coming from each of the vector bosons are reconstructed as a single fat jet ($J$). See, for instance, ref.~\cite{Aoude:2019cmc} for the semileptonic channel, where only one of the vector bosons decays into 2 jets that are reconstructed as a single fat jet.

We consider three categories of events:
\begin{itemize}
  \item Signal: the prediction from our model of the vector resonances for the process~\eqref{mc_process}, with the model parameters set to the corresponding values of one of the BPs of \tabref{tablaBMP}. By construction, this is equal to the SM-EW prediction plus the extra events due to the BSM physics.
  \item SM-EW Background:  SM EW prediction for the process in~\eqref{mc_process}.
  \item SM-QCD Background: SM QCD prediction for the process $pp\to jjJJ$ with 2 light jets being on the VBS region and each of the extra light  jet pairs being reconstructed as one fat jet.
\end{itemize}
Notice that we have not considered other possible backgrounds in this hadronic final state study, like those coming from the already mentioned mixed QCD-EW and top-antitop backgrounds, since they are well below the SM-EW background which we are taking into account together with the dominant and most problematic QCD background. These two considered backgrounds should be sufficient for this study. Notice also that the third category of events from QCD background is quite hard to filter out, as we will describe later.

%%%%%%%%%%%%%%%%%%%%%5

\begin{table}[t!]
\centering
\vspace{.2cm}
%\begin{tabular}{|c p{\mylength}|c p{\mylength}|c p{\mylength}|c p{\mylength}|c p{\mylength}|c p{\mylength}|c p{\mylength}|}
%\begin{tabu} to 1.0 \textwidth {X[c]|X[c]X[c]X[c]X[c]X[c]X[c]X[c]X[c]X[c]}
\begin{tabular}{ c |@{\extracolsep{0.65cm}} c  @{\extracolsep{0.65cm}} c @{\extracolsep{0.65cm}} c@{\extracolsep{0.65cm}} c@{\extracolsep{0.6cm}} c@{\extracolsep{0.65cm}} c@{\extracolsep{0.65cm}} c@{\extracolsep{0.65cm}} c@{\extracolsep{0.65cm}} c }
%\begin{tabular}{ |c|c|c|c|c|c|c|c|c|}
%\rowcolor{gray! 50}
\toprule
\toprule
 &BP1 & BP2  & BP3 & BP1'  & BP2' & BP3' & EW & QCDEW &  $t\bar{t}$ \\[3pt] 
\midrule
$\sigma$ [fb] & 1.57    & 1.46  & 1.44  &  1.8   &  1.55  &  1.51  & 1.43  &  0.71  &   0.11\,(0.24) 
\\[3pt] 
\bottomrule\bottomrule
\end{tabular}
\vspace{0.4cm}
\caption{\small Parton level predictions for the cross sections times the branching ratios of the $W$ bosons to hadrons, $\sigma(pp \to W^+W^-jj)\times({\rm BR}(W\to {\rm hadrons}))^2$ in fb, corresponding to the signal points, BP1, BP2, BP3, BP1', BP2' and, BP3', and for the main backgrounds: SM-EW (EW), mixed SM-QCDEW (QCDEW) and SM top-antitop ($t\bar{t}$). In the $t\bar{t}$ case the decay chain $t \to W b$ has been considered, and a suppression factor of 0.2 (0.3) for each final $b$-jet being misidentified as a light jet $j$  has been applied. All the results are generated with MadGraph at the parton level, summed over the interval $800\,{\rm GeV}< M_{WW}< 3000\,{\rm GeV}$, and the cuts in \eqref{VBScuts} have been applied. 
%The BRs of the hadronic $W$ decays for the two final $W$s have been included.   
}
\label{parton}
\end{table}
%\end{tabular}
%%%%%%%%%%%%%%%%%%%%%%%

The Monte Carlo chain MadGraph~v5~\cite{Alwall:2014hca,Frederix:2018nkq}, Pythia~8~\cite{Sjostrand:2006za} and Delphes~\cite{deFavereau:2013fsa} is used for this analysis. For the jet reconstruction, we use the FastJet library~\cite{Cacciari:2011ma,Cacciari:2005hq} with the anti-$kT$ algorithm~\cite{Cacciari:2008gp}, both integrated inside Delphes. We will also need the boosted objects machinery~\cite{Abdesselam:2010pt,Thaler:2010tr,Thaler:2011gf} integrated in FastJet, for $W$-tagging purposes. 
 
For each event, two lists of reconstructed jets are generated with the anti-$k_T$ jet algorithm, corresponding respectively to the thin (usual) jets, $j$, and to the fat jets, $J$. For the thin-jet one, $R=0.5$ is required, whereas for the fat jet one, $R=0.8$ is required. 
Regarding the cuts on the reconstructed jets, we first apply the following set of initial cuts to the thin jets and to the fat jets, respectively. 
\begin{itemize}
\item[1)] Cuts on the thin jets.

We require 2 thin-jets ($j_1$, $j_2$), not b-tagged~\cite{Cacciari:2011ma,Cacciari:2005hq} that in addition to the detection cuts,  $p_T^{j_1},p_T^{j_2}>20\,{\rm GeV}$, $\Delta R_{jj}>0.4$, verify the VBS cuts,
\begin{gather}
 2 < \lvert\eta_{j_1,j_2}\rvert < 5, \nonumber \\
 \eta_{j_1}\cdot\eta_{j_2} < 0, \nonumber \\
 M_{j_1j_2}>500~{\rm GeV} \, .
\label{VBSthin} 
\end{gather}
\item[2)] Cuts on the fat jets

We  require (at least) 2 fat jets, being $J_1$ and $J_2$ the leading (in the sense of largest $p_T$) and sub-leading fat jet respectively. The following basic cuts are set on the transverse momentum, the mass and the rapidity of each fat jet:
\begin{gather}
 p_T^{J_1}, p_T^{J_2}>200\,{\rm GeV}\,, \nonumber \\
 M_{J_1}, M_{J_2}>20\,{\rm GeV}, \nonumber \\
 \lvert\eta_{J_1}\rvert, \lvert\eta_{J_2}\rvert<2 \, .
\label{fatjetcuts} 
\end{gather}
\end{itemize}
%%%%%%%%%%%%%%%%%%%%%%%%%%%%%%%%%%
\begin{figure}[t!]
\begin{center}
\includegraphics[width=0.49\textwidth]{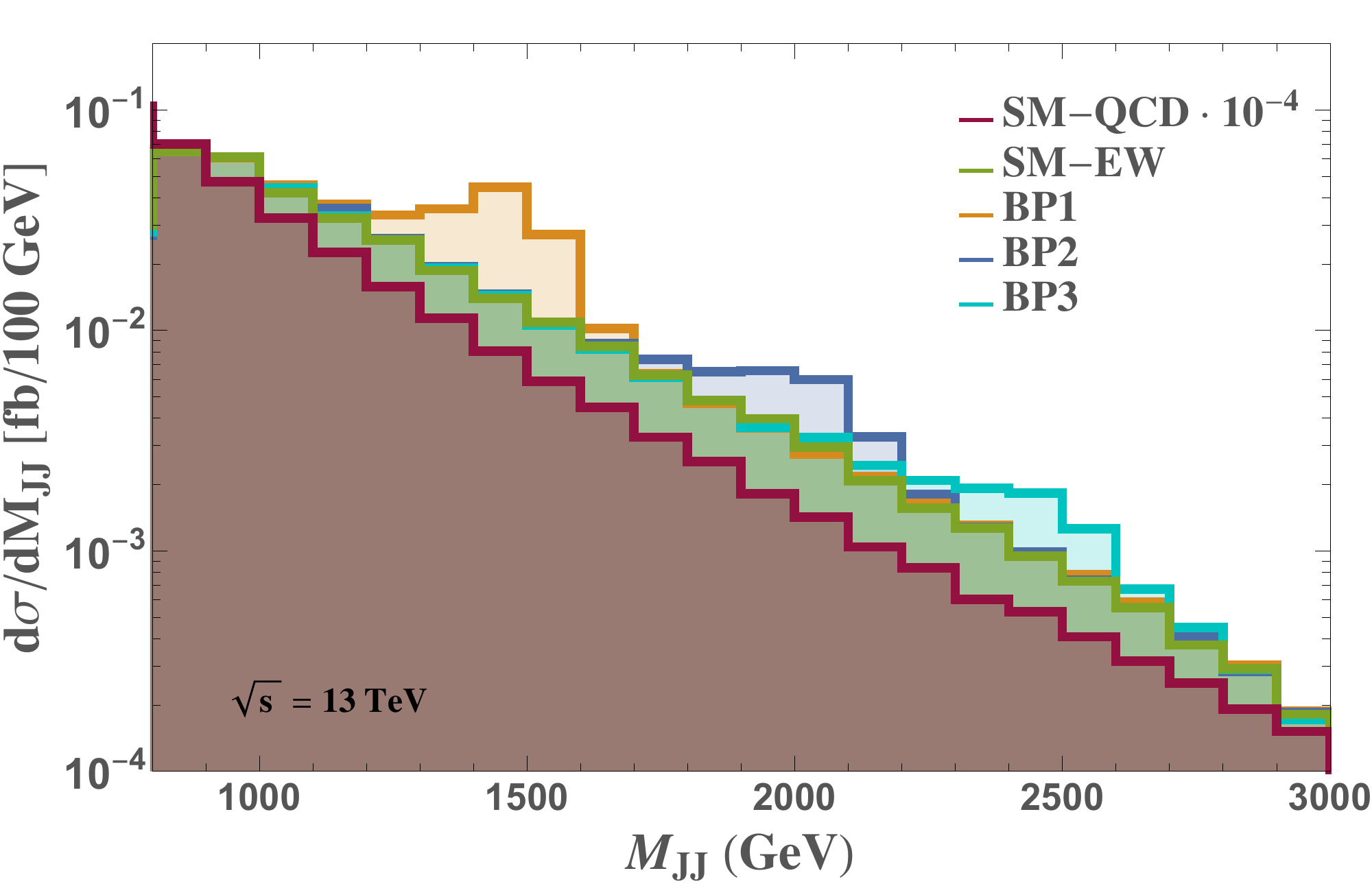}
\includegraphics[width=0.49\textwidth]{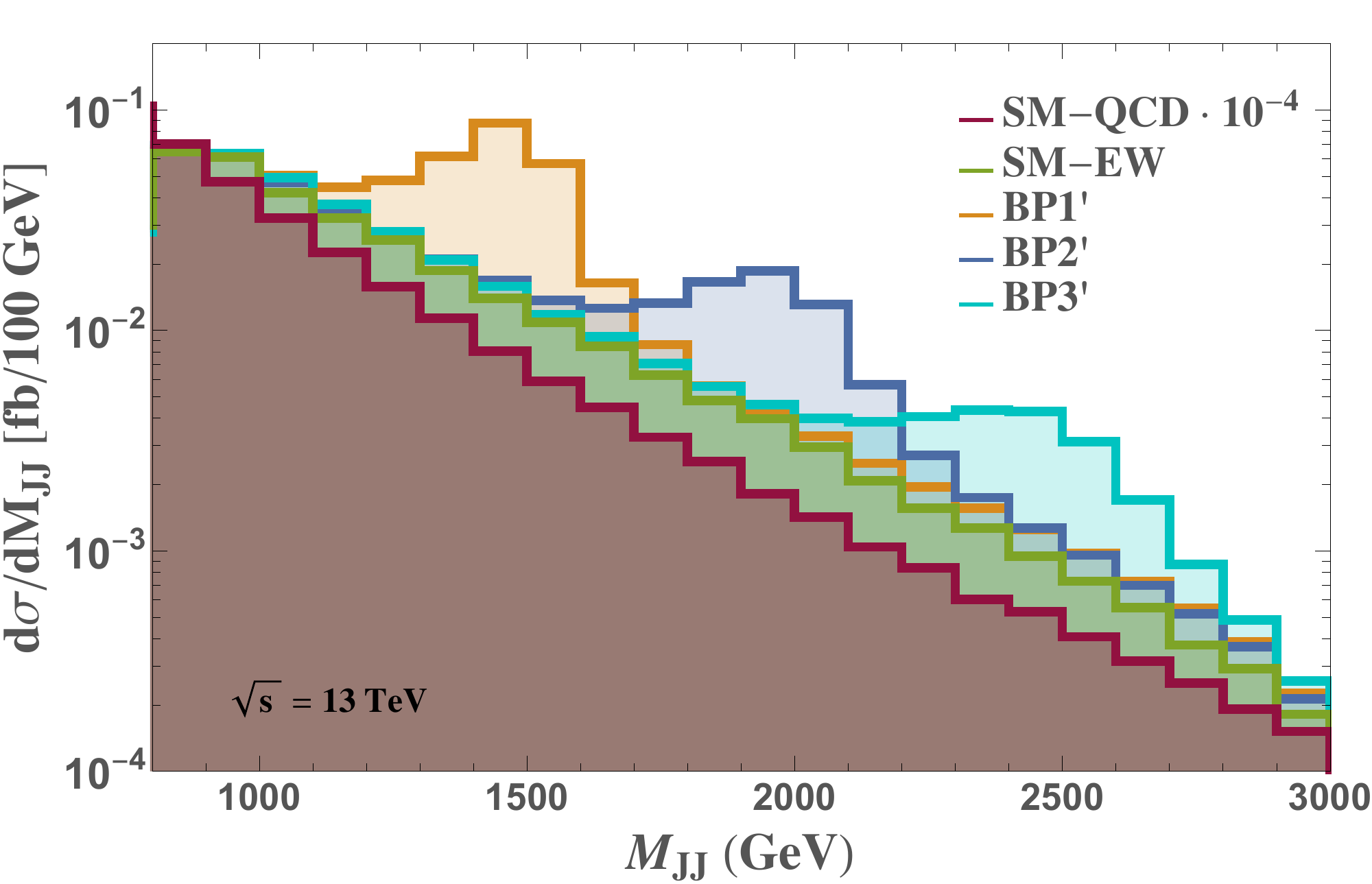}
\caption{Predictions of the cross sections distributions of $JJjj$ events with the invariant mass of the fat jet pair, $M_{JJ}$, after jet reconstruction, using  MadGraph+PYTHIA+DELPHES with the anti-$k_T$ algorithm. The rates for the EW background (SM-EW), the pure QCD background (SM-QCD, scaled down by a factor $10^{-4}$) and the selected signal scenarios for the vector resonances given by the BP's defined in \tabref{tablaBMP} are included. The cuts in \eqref{VBSthin} and \eqref{fatjetcuts} have been applied.}
\label{fig:JJlevel}
\end{center}
\end{figure}
%%%%%%%%%%%%%%%%%%%%%%%%%%%%%%%%%%%%%%%%%%
If the event is correctly identified, the $j_1$ and $j_2$ VBS jets will be the 2 reconstructed jets $jj$ coming from the $pp\to W^+W^-jj$ VBS event. The $J_1$ and $J_2$ fat jets will then correspond, directly, to the reconstructed vector bosons $W^\pm\to J(jj)$. By means of 4-momenta conservation, the masses $M_{J_1}$ and $M_{J_2}$ and the total invariant mass $M_{JJ} = (p_{J_1} + p_{J_2} )^2$ of the reconstructed fat jets are identified with those of the original vector bosons coming from the VBS event. Notice that because of the usage of fully hadronic events no information is lost, in contrast to the cases where there are neutrinos in the final state. Hence, the component of momenta parallel to the beamline can be reconstructed. Note also that the requirement $2 < \lvert\eta_{j_1,j_2}\rvert < 5$ for the VBS thin-jets and $\lvert\eta_{J_1,J_2}\rvert<2$ for the reconstructed fat jets coming from the vector bosons means that both objects can be (in principle) easily classified by means of $\eta$ variable since they belong to mutually excluding regions.

Actually, a fat jet coming from a vector boson can also appear in the thin-jet reconstruction list, both as a single jet or even as 2 jets with a small separation. In the first case, a very boosted vector boson decaying into 2 jets would be reconstructed as a fat jet with a relatively small radius, that could be also reconstructed as a single thin jet. In the second case, the thin jet algorithm has been able to reconstruct the decay products ($jj$) of the vector boson, that is likely to happen for not-so-boosted vector bosons. Note that these additional thin jets cannot be confused with VBS jets because of the different ranges of the variable $\eta_j$.

It is also possible that additional jets collinear with those coming from the hard scattering event are produced by final state radiation. These additional jets could be reconstructed as additional thin jets, or could lead to a thin jet being also reconstructed as a fat jet. The fat jet constituents would be those coming from the additional radiation process.
%%%%%%%%%%%%%%%%%%%%%5
\begin{table}[h!]
\centering
\vspace{.2cm}
%\begin{tabular}{|c p{\mylength}|c p{\mylength}|c p{\mylength}|c p{\mylength}|c p{\mylength}|c p{\mylength}|c p{\mylength}|}
%\begin{tabu} to 1.0 \textwidth {X[c]X[c]X[c]X[c]X[c]X[c]X[c]X[c]X[c]}
\begin{tabular}{ c |@{\extracolsep{0.8cm}} c  @{\extracolsep{0.8cm}} c @{\extracolsep{0.8cm}} c@{\extracolsep{0.8cm}} c@{\extracolsep{0.8cm}} c@{\extracolsep{0.8cm}} c@{\extracolsep{0.8cm}} c@{\extracolsep{0.8cm}} c@{\extracolsep{0.85cm}} c }
%\begin{tabular}{ |c|c|c|c|c|c|c|c|c|}
%\rowcolor{gray! 50}
\toprule
\toprule
 &BP1 & BP2  & BP3 & BP1'  & BP2' & BP3' & EW & QCD \\[3pt]
\midrule
$\sigma$[fb]&  0.384     & $ 0.322  $ & $ 0.312  $ & $  0.526   $ & $  0.380  $ & $  0.348  $ & $  0.304  $ & $  2310 $ 
\\[3pt]
\bottomrule\bottomrule
\end{tabular}
\vspace{0.4cm}
\caption{\small Predictions after jets reconstruction for the cross sections, $\sigma(pp\to JJjj)$ in fb, corresponding to the signal points, BP1, BP2, BP3, BP1', BP2' and, BP3', and for the main backgrounds: SM-EW (EW) and SM-QCD (QCD).  The results are generated with MadGraph+PYTHIA+DELPHES, summed over the interval $800\,{\rm GeV} < M_{JJ} < 3000\,{\rm GeV}$, and the cuts in \eqref{VBSthin} and \eqref{fatjetcuts} have been applied.   
}
\label{JJ}
\end{table}
%\end{tabular}
%%%%%%%%%%%%%%%%%%%%%%%

With all the above considerations taken into account, we make our predictions for the three specified event categories  and find the results summarized in  \figref{fig:JJlevel} and  \tabref{JJ}. \figref{fig:JJlevel} shows the distributions in the invariant mass of the $JJ$ pair and \tabref{JJ} summarizes the total cross sections in the invariant mass region of the $JJ$ pair of our interest here, i.e., summing events over the interval $800\,{\rm GeV}<M_{JJ}<3000\,{\rm GeV}$. The main conclusions we learn from these first results are the following: first, we see clearly in \figref{fig:JJlevel} that the vector resonances still emerge over the EW background, although with wider peaks  than in the previous results at the partonic level, due to the  `typical energy loss' in the jet reconstruction process. Second, when we compare the $WWjj$ parton level rates in \tabref{parton} with the $JJjj$ rates in  \tabref{JJ} we see that the ratios $JJjj/WWjj$ for the EW processes, i.e., the EW background and the signal BP's, are in the interval $(0.2,0.3)$ which can be interpreted as coming from an efficiency in each $W$ tagging from each fat jet in the range $(0.45,0.55)$, and this is in agreement with previous estimates of this efficiency (see, for instance refs.~\cite{Khachatryan:2014hpa,Aad:2015rpa,Aad:2015owa,Heinrich:2014kza}). We also learn from these results of the total $JJjj$ rates that the dangerous QCD background overwhelms both the signal and the EW background by a factor of $10^3-10^4$. Concretely, the total cross section integrated over the interval $800\,{\rm GeV}< M_{JJ} < 3000\,{\rm GeV}$ for the QCD background is, according to our result in \tabref{JJ}, 4392 times larger than our largest signal (the BP1'). This fact is really challenging to deal with. Therefore, in order to improve the signal to background ratios a more refined analysis benefiting from the fat jet features is needed. 
%%%%%%%%%%%%%%%%%%%%%%%%%%%%%%%%%%
\begin{figure}[t!]
\begin{center}
\includegraphics[width=0.49\textwidth]{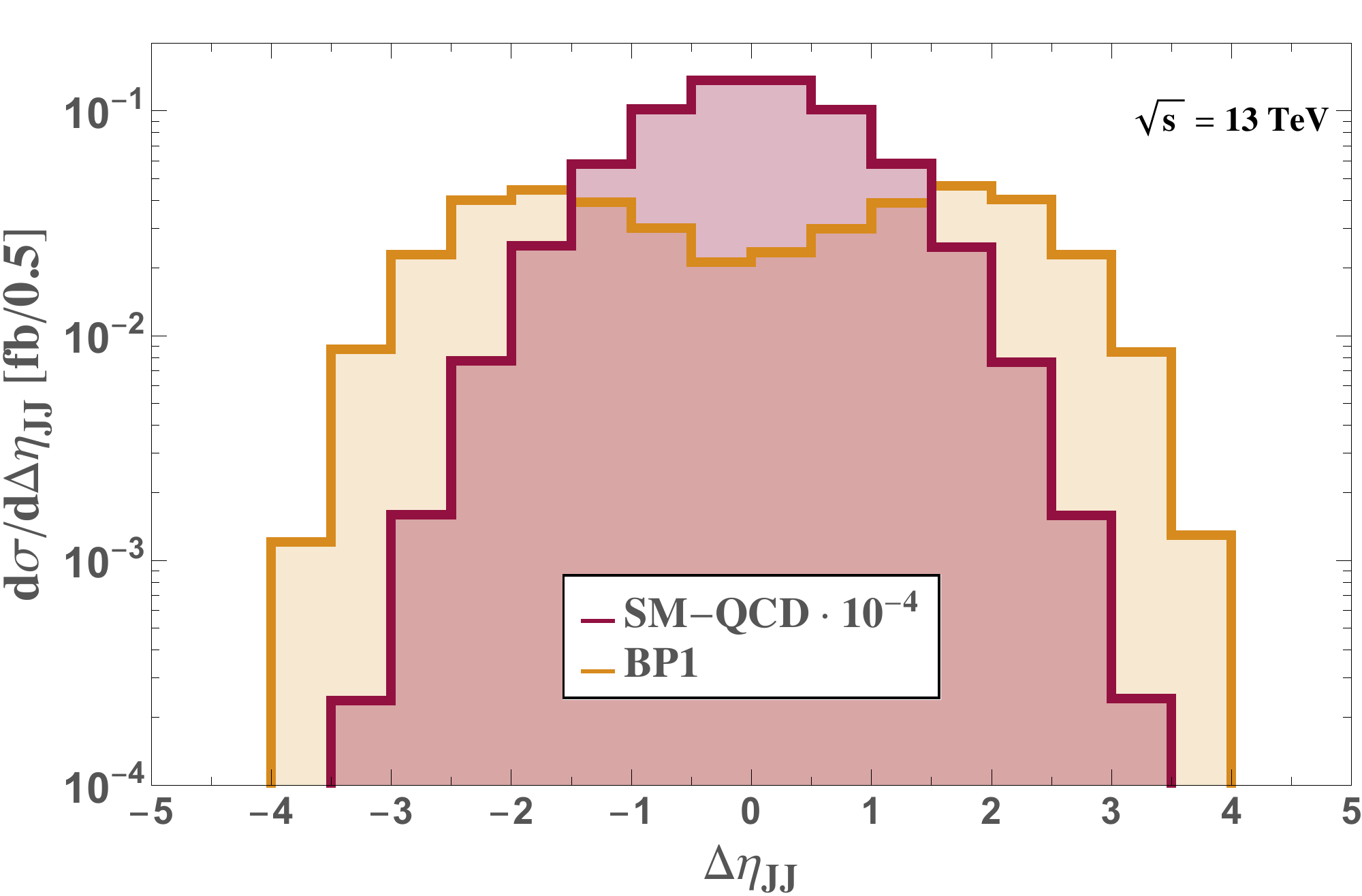}
\includegraphics[width=0.49\textwidth]{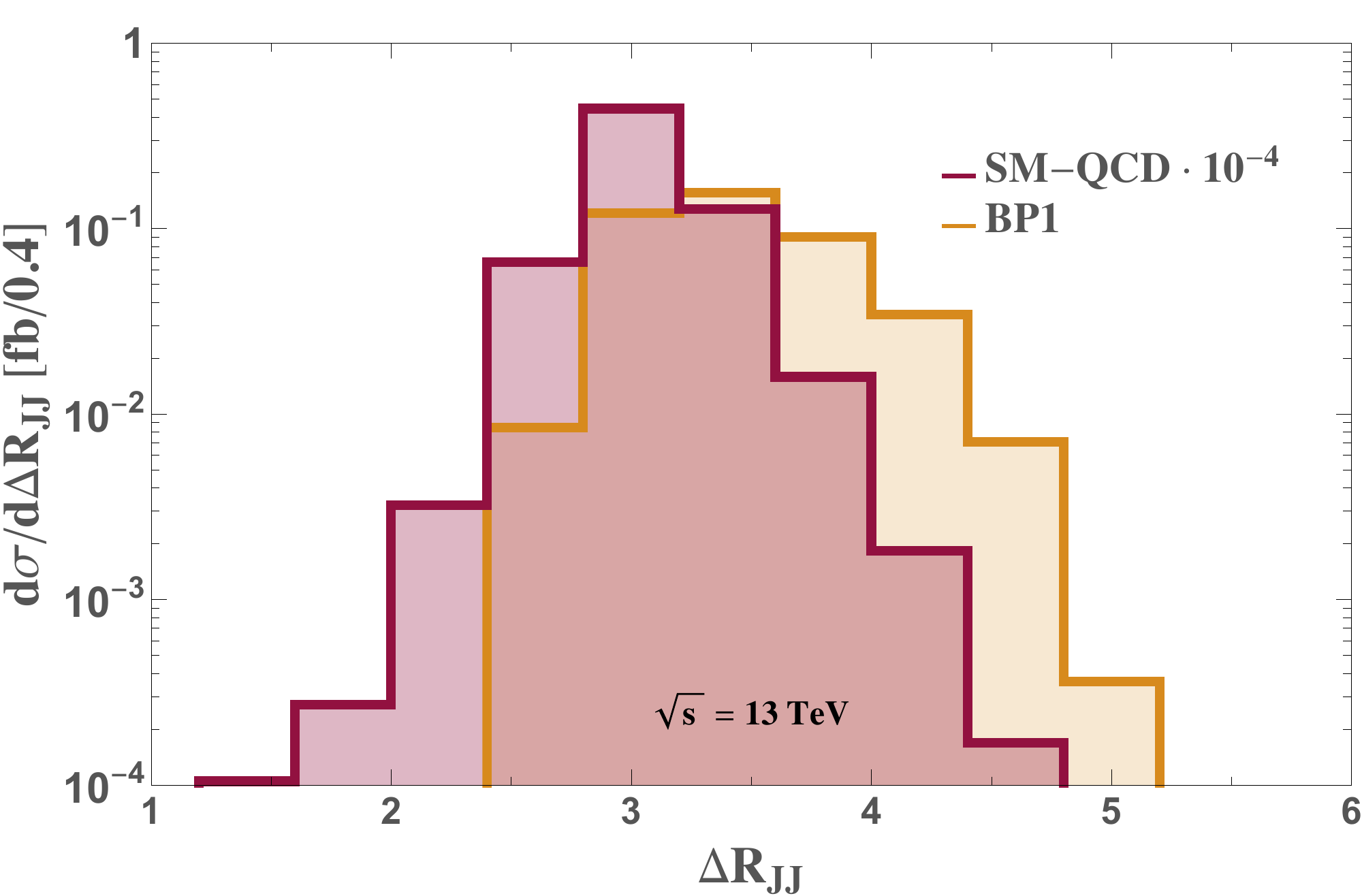}
\includegraphics[width=0.49\textwidth]{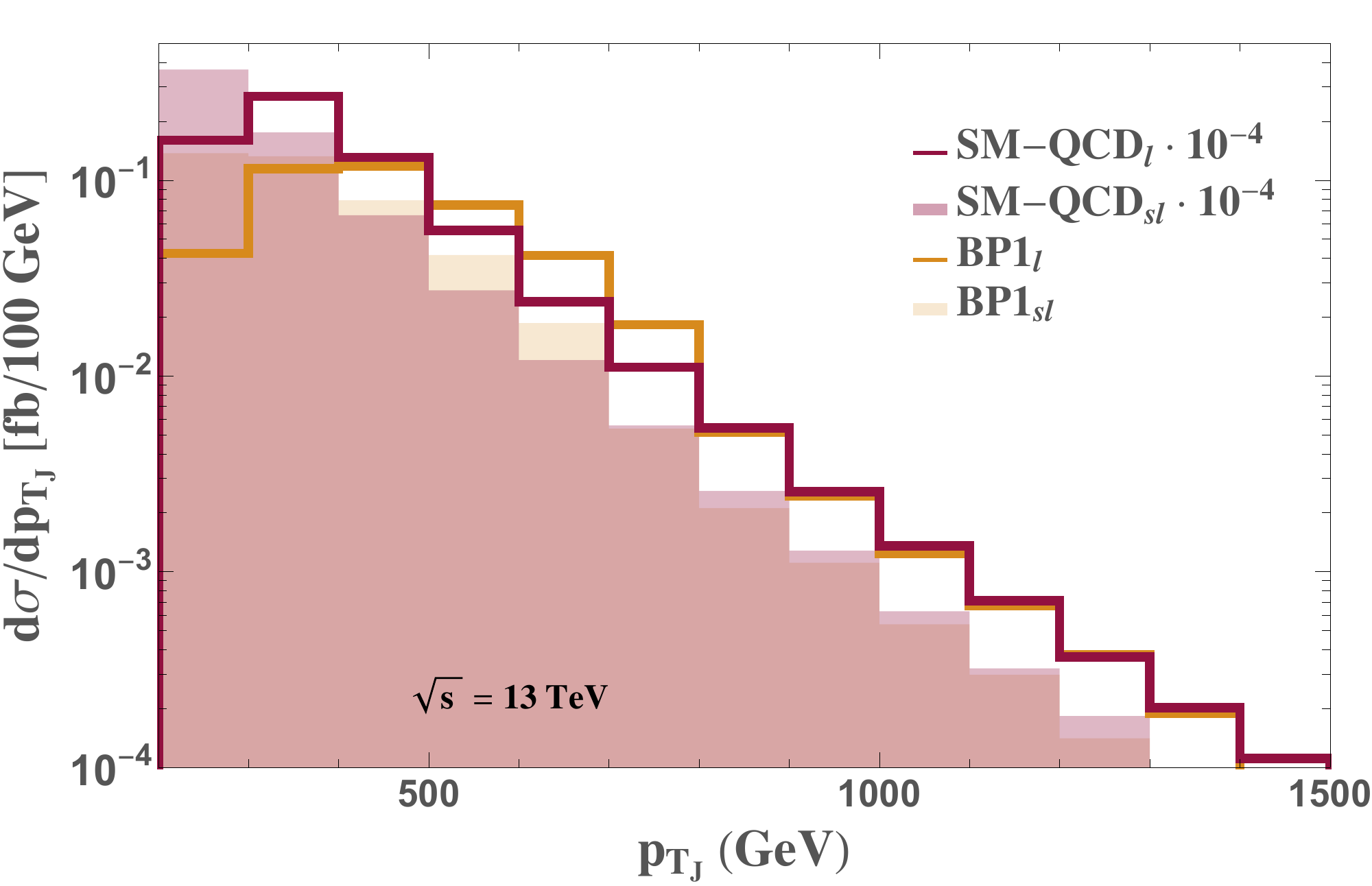}
\includegraphics[width=0.49\textwidth]{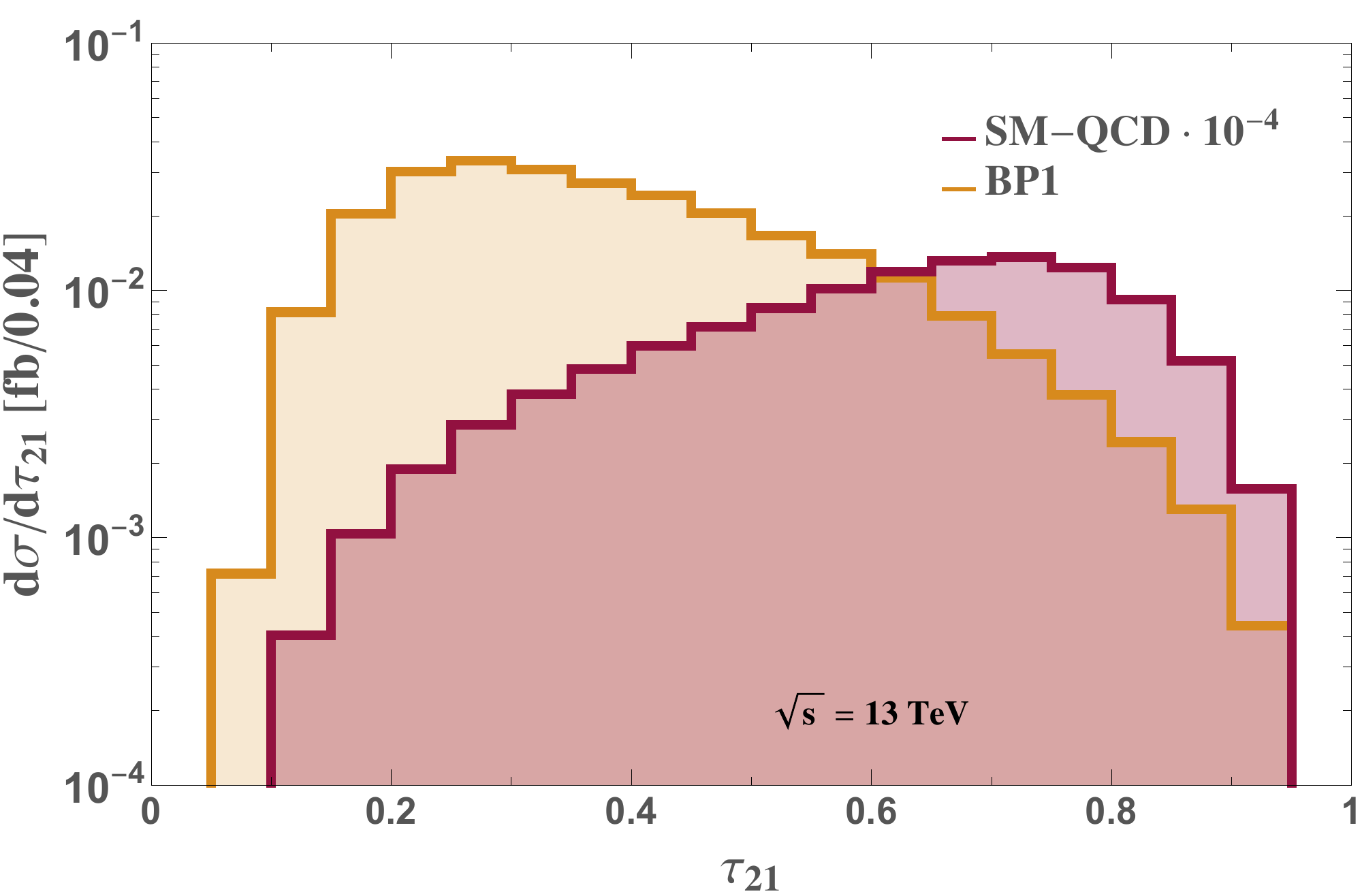}
\caption{Predictions, after jets reconstruction, of the cross sections distributions with the fat jet variables $\Delta\eta_{JJ}$ (upper left plot), $\Delta R_{JJ}$ (upper right plot), $P_{T_J}$ (lower left plot) and $\tau_{21}$ (lower right plot) for the QCD background (SM-QCD, scaled down by a factor $10^{-4}$) and for the signal in the  scenario BP1. The cuts in \eqref{VBSthin} and \eqref{fatjetcuts} have been applied.}
\label{fig:Jfeatures}
\end{center}
\end{figure}
%%%%%%%%%%%%%%%%%%%%%%%%%%%%%%%%%%%%%%%%%%

We have investigated further into more specific characteristics of the produced fat jets, analyzing in more detail the events for both signal and QCD background in terms of the following  fat jet variables and their optimal cuts: $M_{J_1}$, $M_{J_2}$, $p_T^{J_1}$, $p_T^{J_2}$, $\Delta\eta_{JJ} = \eta_{J_1} - \eta_{J_2}$, 
$\Delta R_{JJ}=\sqrt{(\Delta (\Phi_{JJ})^ 2+(\Delta\eta_{JJ})^2}$, and $\tau_{21}=\tau_2/\tau_1$. 
In particular, the latter variable $\tau_{21}$~\cite{Thaler:2010tr} seems to be a very good discriminant for boosted objects studied via fat jets. In fact, it has been already used for $W$-tagging purposes in the case of the semileptonic decay channel~\cite{Aoude:2019cmc} and in the recent ATLAS study~\cite{Aad:2019fbh}.

First of all, it is obvious that a cut on the mass variables like $M_J$ and $M_{JJ}$ by restricting them to windows around, respectively,  the $M_W$ mass and the corresponding resonance mass $M_{V^0}$ of the studied BP signal,  will improve considerably the signal to background ratio. The only problem setting  these mass windows is that if they are too narrow we may loose too much signal and end up lacking statistics for the analysis. In this concern, it should be noticed that we are talking about emergent peaks over the EW background containing in the best case $(\sigma_{\rm S}-\sigma_{\rm EW}) \sim 0.222\,{\rm fb}$, therefore, a total number of events of at most 67 for an integrated luminosity of $L = 300 \, {\rm fb}^{-1}$, and around 670 for $L = 3000\,{\rm fb}^{-1}$. For this reason, we have focused our more refined study of the events to the highest luminosity option.   
 
Regarding the remaining fat jet variables, we present our results for the distributions of the signal and the QCD-background with respect to  $\Delta\eta_{JJ}$, $\Delta R_{JJ}$, $p_T^{J_i}$ and $\tau_{21}$ in \figref{fig:Jfeatures}. For the signal we have selected the BP1 case, as an example. From these figures we learn that the two fat jets from the BP1 signal tend to be more separated, both in $\Delta\eta_{JJ}$ and in $\Delta R_{JJ}$, than in the QCD case. Also the transverse momenta $P_{T_J}$ of the leading (l) and subleading fat jets (sl), tend to be larger in the signal case than in QCD one. Finally, we find that the $\tau_{21}$ variable which tests the correctness of the hypothesis of having a fat jet being composed of two light jets (with $\tau_{21}$ close to 1 meaning that this hypothesis is incorrect), is one of the best discriminants in our case. For instance, we have checked explicitly that applying a cut of $\tau_{21}<0.3$ together with $60\,{\rm GeV}< M_J < 100\,{\rm GeV}$, in the large invariant mass selected interval $1000\,{\rm GeV}<M_{JJ}<3000\,{\rm GeV}$, reduces the QCD background by a factor of $2.4 \times 10^{-5}$ whereas the BP1 (BP1') signal is reduced by a milder factor of  $6.3 \times 10^{-2} (7.8 \times 10^{-2})$. Thus, this $\tau_{21}$ variable together with $M_J$ is very efficient in reducing the QCD background to a controllable level. Again the only problem is the low statistics of the signal when imposing tight cuts, specially for the heavier resonances.

Finally, in order to make a more systematic exploration looking for the best set of cuts on the fat jet variables, we have done a full survey considering all the possible combinations of cut choices, including four options for each variable cut. These are set in addition to the basic cuts in \eqref{VBSthin} and \eqref{fatjetcuts}. The following options have been considered:
\begin{itemize}
\item[] $M_J\,({\rm GeV})$ in the interval (50,110), (60,100), (70,95), or no cut. $J$ refers to both fat jets. 
\item[]  $p_T^J\,({\rm GeV})$ minimum: 200, 300, 400, 600. $J$ refers to both fat jets. 
\item[] $\lvert\Delta \eta_{J_1J_2}\rvert$ minimum: 0.5, 1.0, 1.5, or no cut
\item[] $\Delta R_{J_1J_2}$ in the interval (2,5), (2.5,4.5), (3,4), or no cut. 
\item[] $\tau_{21}(J)$ in the interval (0.1,0.4), (0.1,0.35), (0.1,0.3), or no cut. $J$ refers to both fat jets. 
\end{itemize}

The considered windows in $M_{JJ}$ are fixed correspondingly for each BP scenario to the interval centered at approximately the resonance mass $m_{V^0}\pm 250$~GeV. In addition, in our search of the optimal cuts we also allow for events with up to 4 extra thin jets, besides to the two VBS thin jets, with $\Delta R_{jJ}<0.8$ (angular distance between the non-VBS thin jet $j$ and the closest fat jet $J$).  Regarding the fat jets, we require a minimum of two reconstructed fat jets and a maximum of four, and the variable $M_{JJ}$ is the reconstructed invariant mass of the two leading fat jets.  In all this study of optimal cuts we have used the highest luminosity option for the LHC of $3000\,{\rm fb}^{-1}$. Notice also that due to the fact that it is not possible to disentangle the $W^+W^-jj$ case from the $W^+W^+jj$ one if one uses $W$ tagging by means of fat jets, we consider in this last analysis both final states contributing in both EW background and signal predictions. In fact, also the $W^-W^-jj$ case would contribute to our final $JJjj$ events but we have checked that it is much smaller than the other two cases. Specifically, for the EW background, we have found the following hierarchy in their corresponding rates at the parton level: $\sigma(W^+W^-jj)=2.4 \, \sigma(W^+W^+jj)=11.2\,\sigma(W^-W^-jj)$. Thus, in the following, we have neglected the contribution from $W^-W^-jj$ and $W^+W^+jj$ in both signal and EW background rates. 

%%%%%%%%%%%%%%%%%%%%%%%

\begin{table}[t!]
\centering
\vspace{.2cm}
%\begin{tabular}{|c p{\mylength}|c p{\mylength}|c p{\mylength}|c p{\mylength}|c p{\mylength}|c p{\mylength}|c p{\mylength}|}
\begin{tabu} to \textwidth {X[c]|X[c]X[c]X[c]X[c]X[c]X[c]}
%\begin{tabular}{ c|ccccccc }
%\rowcolor{gray! 50}
\toprule
\toprule
& {\footnotesize {\bf BP1'}} & {\footnotesize {\bf BP2'}} &  {\footnotesize {\bf BP3'}} & {\footnotesize {\bf BP1}} & {\footnotesize {\bf BP2}} & {\footnotesize {\bf BP3}} 
\\
\midrule
$\sigma^{\rm stat}$              & 11.7        & 4.1             & 2.3       & 4.6        & 0.77     & 0.52
\\[3pt]                                                                                                             
$\sigma_{\rm EW}^{\rm stat}$               & 18.8         & 8.3             & 5.8       & 7.3        & 1.6      & 1.3
\\[3pt]                                                                                                             
$N_{\rm S}$                           & 103         & 17.2               & 9.7       & 51.9        & 5.7      & 3.6
\\[3pt]                                                                                                             
$N_{\rm EW}$                    & 19.5        & 2.9                & 1.8        & 19.5        & 2.9      & 1.8
\\[3pt]                                                                                                             
$N_{\rm QCD}$                   & 30.8         & 9.5             & 9.8        & 30.8         & 9.5      & 9.8
\\[3pt]                                                                                                             
\midrule                                                                                                            
$p_T^{J_1}$(GeV)                & $>200$           & $>200$      & $>200$      & $>200$      & $>200$    & $>200$     
\\[3pt]                                                                                                             
$p_T^{J_2}$(GeV)                & $>200$           & $>200$      & $>200$      & $>200$      & $>200$    & $>200$  
\\[3pt]                                                                                                             
$\tau_{21}(J_1)$                & $0.1-0.3$    & $0.1-0.3$   & $0.1-0.4$   & $0.1-0.3$   & $0.1-0.3$ & $0.1-0.4$
\\[3pt]
$\tau_{21}(J_2)$                & $0.1-0.4$     & $0.1-0.3$   & $0.1-0.4$   & $0.1-0.4$   & $0.1-0.3$ & $0.1-0.4$ 
\\[3pt]                                                                                                             
$M_{J_1}$(GeV)                   & $70-95$    & $70-95$      & $70-95$    & $70-95$    & $70-95$  & $70-95$  
\\[3pt]   
$M_{J_2}$(GeV)                   & $70-95$    & $50-110$      & $50-110$    & $70-95$    & $50-110$  & $50-110$  
\\[3pt]                                                                                                          
$\lvert\Delta \eta_{J_1J_2}\rvert_{\rm min}$ & no cut         & no cut        & no cut         & no cut       & no cut  & no cut 
\\[3pt]  
$\Delta R_{J_1J_2}$ & no cut          & no cut        & no cut         & no cut         & no cut  & no cut  
\\[3pt]    
$M_{JJ}$(GeV)      & $1500 \pm 250$ & $2000 \pm 250$  & $2500 \pm 250$ & $1500\pm 250$ & $2000\pm 250$ & $2500\pm 250$ 
\\[3pt]
\bottomrule\bottomrule
\end{tabu}
%\end{tabular}
\caption{\small Results for the optimal cuts and the number of predicted events for the various signal scenarios ($N_{\rm S}$) given by the BPs and for the SM backgrounds, EW ($N_{\rm EW}$) and QCD ($N_{\rm QCD}$), as well as their associated statistical significances defined in \eqref{stat}. Here $J_{1,2}$ are the leading and subleading fat jets respectively. The integrated luminosity is fixed to  $L=3000\,{\rm fb}^{-1}$.}
\label{tablaSIGMAS}
\end{table}

The results of our survey for the optimal cuts are collected in \tabref{tablaSIGMAS}. We define our optimal cuts as those that lead to the best statistical signal significance, defined as:
\begin{align}
  \sigma^{\rm stat} &= \frac{N_{\rm S}-N_{\rm EW}}{\sqrt{N_{\rm EW}+N_{\rm QCD}}} \, ,\\
  \sigma_{\rm EW}^{\rm stat}  &= \frac{N_{\rm S}-N_{\rm EW}}{\sqrt{N_{\rm EW}}},
\label{stat}
\end{align}
where $N_{\rm S}$,  $N_{\rm EW}$ and  $N_{\rm QCD}$ refer to the number of events of the signal, EW background and QCD background, respectively, for a  luminosity of $L=3000~{\rm fb}^{-1}$. Due to the clear dominance of the QCD background over the EW background and our incapability  of separating both background sources, the most realistic and determining value should be that of the $\sigma^{\rm stat}$ above.

In \tabref{tablaSIGMAS} we summarize our main findings and shows that the best significances are found for the resonances with mass close to $1500$~GeV, especially BP1', and next for BP2' with a mass around $2000$~GeV. The expectations for the resonances with a heavy mass close to 3000 GeV are considerably less appealing. Notice that we have also included $ \sigma_{\rm EW}^{\rm stat}$ in our results in \tabref{tablaSIGMAS} mainly to motivate for a future more sophisticated analysis that could find out a more efficient way to suppress the difficult QCD background. In that hypothetical case the statistical significances of the signal would improve considerably, as can it be learnt  from the appealing larger values of $ \sigma_{\rm EW}^{\rm stat}$ in this Table.  

In the performed analysis leading to our final results summarised in \tabref{tablaSIGMAS} we have also learnt some interesting features regarding the comparative performance of the various cuts on the studied variables $p_T^{J_{1,2}}$, $\tau_{21}(J_{1,2})$, $M_{J_{1,2}}$, $\lvert\Delta\eta_{J_1J_2}\rvert_{\rm min}$, $\Delta R_{J_1J_2}$, and $M_{JJ}$, which we believe are worth to comment. Firstly, as common features to all the studied points, we find that the best cut on $p_T^{J_{1,2}}$ for both fat jets is $p_T^{J_{1,2}}>200$~GeV. Applying a stronger cut on $p_T^{J_{1,2}}$, like the ones considered in our analysis of $300$, $400$ and $600$~GeV, leads generally to a smaller statistical significance of the signal, basically because of the lack of statistics for the signal. It is also common to all points, that $\sigma^{\rm stat}$ is not very sensitive to the considered cuts on $\lvert\Delta \eta_{J_1J_2}\rvert_{\rm min}$ and $\Delta R_{J_1J_2}$, and, therefore, the best option again in order not to loose much signal is the 'no cut' option in both of these variables. For instance, for BP1', varying $\lvert\Delta \eta_{J_1J_2}\rvert_{\rm min}$ from our best option, 'no cut', to 1 changes the prediction of $\sigma^{\rm stat}$ from the largest value in \tabref{tablaSIGMAS} of 11.7 to 10.5. A similar feature is found for the other points. Therefore, the most efficient cuts in reducing the QCD background and leading to the largest $\sigma^{\rm stat}$ for all the benchmark points are those applied on $M_{J_{1,2}}$ and on $\tau_{21}(J_{1,2})$. Since we have done our exploration of cuts for $\tau_{21}(J_{1,2})$ by starting already with quite optimised and similar intervals, we do not find much differences in their corresponding predictions of $\sigma^{\rm stat}$ for all the points. For instance, for BP1', varying $\tau_{21}(J_{1,2})$ from the combination in \tabref{tablaSIGMAS}, respectively,  of ((0.1,0.3), (0.1,0.4)) to ((0.1,0.3),(0.1,0.3)) changes $\sigma^{\rm stat}$ from 11.7 to  11.4. The other combination considered, ((0.1,0.4),(0.1,0.4)) leads to a slightly lower $\sigma^{\rm stat}$ of 8.1. Once again a similar conclusion applies to  the other points. Secondly, the performance of the cuts explored on the variables $M_{J_{1,2}}$ deserves some devoted comments. These are very relevant variables in the definition of our signal, since they quantify the accuracy in the identification/reconstruction of the two final gauge bosons from the two selected fat jets. Therefore, a priori, the more precision in the determination of $M_{J_{1,2}}$ the better the expected statistical significance of the signal. However, considering too narrow intervals in these variables does not always lead to the best  $\sigma^{\rm stat}$, again because of the lack of statistics for the signal, specially for the heavier resonances. The best option that we have found, as shown in \tabref{tablaSIGMAS}, is: $M_{J_1}$(GeV) in (70,95), $M_{J_2}$(GeV) in (70,95) for BP1' and BP1; and $M_{J_1}$(GeV) in (70,95), $M_{J_2}$(GeV) in (50,110) for BP2', BP2, BP3' and BP3. Changing these to other options leads to lower $\sigma^{\rm stat}$. For instance, for BP1', considering $M_{J_1}$(GeV) in (70,95), $M_{J_2}$(GeV) in (60,100), changes  $\sigma^{\rm stat}$ from our best value of 11.7 to 10.6; and considering $M_{J_1}$(GeV) in (70,95), $M_{J_2}$(GeV) in (50,110) changes it to 9.6. Similarly, for BP2', considering $M_{J_1}$(GeV) in (70,95), $M_{J_2}$(GeV) in (60,100) changes $\sigma^{\rm stat}$ from the best value $4.1$ to $3.9$, and considering  both $M_{J_{1,2}}$(GeV) in (70,95), changes it to 3.0. Finally, it is also interesting to comment on what happens if the search is not devoted to the starting optimised windows in $M_{JJ}$, as indicated in \tabref{tablaSIGMAS}, corresponding to the explored intervals centered  at the mass of each resonance. After all, when one compares with data there is not a preferred interval in $M_{JJ}$ where to look at a priori. Thus, it is also interesting to explore the optimal combination of cuts for the overall best performance across all the points in the full explored region of $M_{JJ}$(GeV) in (1000,3000). In this sense, we have checked that the set of cuts specified in \tabref{tablaSIGMAS}, other than $M_{JJ}$,  lead to the best options for all the points. But, as expected, the statistical significance of the signal decreases significantly for all the points compared to the case in which the $M_{JJ}$ window is optimized for each resonance mass.  Specifically, for $M_{JJ}$(GeV) in (1000,3000), we find that $\sigma^{\rm stat}$ for BP1', BP2', BP3', BP1, BP2 and BP3, change our best values in \tabref{tablaSIGMAS} to 4.8, 1.1, 0.5, 1.8, 0.1 and 0.03 respectively. Therefore, in that case, only BP1' could be observed.

\section{Conclusions}
\label{conclusions}

In this work we have extended our previous UFO model first applied to the study of the vector resonances emerging in the VBS channel $W^+Z\to W^+Z$ in~\cite{Delgado:2017cls} to the different VBS channel $W^+W^-\to W^+W^-$. The model was built in~\cite{Delgado:2017cls} from the EChL, the IAM and the Proca Lagrangian and predicts a triplet of dynamical vector resonances, $V^0$, $V^\pm$ with the relevant physical properties, like the mass, the width and the couplings to EW gauge bosons given in terms of the EChL parameters. The charged vector resonance $V^+$ was relevant for $W^+Z\to W^+Z$ since it can propagate in the s-channel of this process, leading to emergent peaks in the cross section. The neutral resonance  $V^0$ can propagate instead in the s-channel of  the $W^+W^-\to W^+W^-$ process, therefore leading to a resonant peak in the corresponding cross section. Our study here of these resonances at the LHC by means of the process $pp \to WW jj$ is therefore complementary to the previous study in~\cite{Delgado:2017cls}. Furthermore, in contrast to~\cite{Delgado:2017cls} where the focus was set on the fully leptonic channel of the final EW gauge bosons, we devote here our full work to the difficult but very interesting task of disentangling the vector resonance in $pp \to WW jj$ events using instead the $W$ gauge boson hadronic decays. 

We have performed a detailed study of the final hadronic states in $pp \to JJjj$ events at the LHC with two thin jets having the typical VBS topology and using the two fat jets for the tagging of the two final $W$ bosons. We have worked in the scenario where the two thin jets from the $W$ decay are reconstructed as a single, large radius fat jet $J$. After the devoted reconstruction of the final jets we have put all our focus on searching for techniques that can reduce efficiently the most challenging background coming from QCD. We have found that the specific variables $\tau_{21}(J)$, $M_J$ and $M_{JJ}$ are extremely helpful in this concern. 
We have performed a survey on the expected rates for both the signal and the background, as well as for the corresponding statistical significances of the associated BSM signal from the vector resonances, for the six selected benchmark points of \tabref{tablaBMP}. The main results have been summarized in \tabref{tablaSIGMAS} where the optimal cuts that we found are also shown. 

Our main conclusions are that, with these optimal cuts, only the BP1' largely overcomes the $5\sigma$ of statistical significance at the LHC with a luminosity of $L=3000\,{\rm fb}^{-1}$, provided that the MadGraph+Pythia8+Delphes model of the QCD background is accurate and no other relevant backgrounds exist. The other point with the resonance mass close to 1500 GeV, BP1, reaches a statistical significance slightly below~5. The BP2' with a resonance mass close to $2000$~GeV reaches a statistical significance of $4.1\sigma$. For BP3' with mass close to $2500$~GeV we find a maximum in $\sigma^{\rm stat}$ of~2.3. The BP2 and BP3, with a statistical significance below~1, are extremely challenging to observe. And in the hypothetical situation that the QCD background were efficiently suppressed, all the points, except BP2 and BP3, could be observed. This is similar to the situation in our previous study on the fully leptonic $pp\to W^+Zjj$ (ref.~\cite{Delgado:2017cls}).

Hence, the fully hadronic channel could be studied with $L=3000\,{\rm fb}^{-1}$, but the QCD background is a great challenge. It would be necessary to perform a more detailed study of the SM QCD background, since this is the main issue, and, of course, to improve the signal vs. background ratio. Some more sophisticated techniques like deep learning, boost decision trees and others could be used to find better optimal cuts, in a per-event basis, and dealing with hidden correlations between the different collider variables, particularly those characterising the fat jets. According to \tabref{tablaSIGMAS}, BP1', BP2' and BP1 would be clearly detectable at $L=3000\,{\rm fb}^{-1}$.

\section*{Acknowledgments}
This work is supported by the European Union through the ITN ELUSIVES H2020-MSCA-ITN-2015//674896 and the RISE INVISIBLESPLUS H2020-MSCA-RISE-2015//690575, by the CICYT through the projects FPA2016-78645-P, by the Spanish Consolider-Ingenio 2010 Programme CPAN (CSD2007-00042) and by the Spanish MINECO's ``Centro de Excelencia Severo Ochoa''  Programme under grant SEV-2016-0597.  
R.L.D. is financially supported by the Ram\'on Areces Foundation.
%Official C2PAP support and usage of T30 cluster
We acknowledge the support by the DFG Cluster of Excellence ``Origin and Structure of the Universe''. The simulations have been carried out on the computing facilities of the Computational Center for Particle and Astrophysics (C2PAP) and on a local cluster at T30 department at Technische Universit\"at M\"unchen (TUM). We also wish to thank the Theory Department at CERN for allowing us to have access to the computing facilities at CERN which have been of great help for the heavy computations and simulations performed in this work. 

\bibliographystyle{JHEP}
\bibliography{DGH}

\end{document}